\PassOptionsToPackage{hyphens}{url}
\documentclass[runningheads]{llncs}

\usepackage{amsmath}
\usepackage{amsfonts}
\usepackage{amssymb}
\usepackage{booktabs}
\usepackage{graphicx}
\usepackage[colorlinks]{hyperref}
\usepackage[switch]{lineno}
\usepackage[dvipsnames]{xcolor}
\usepackage{cancel}

\graphicspath{{figures/}}

\newcommand{\appendixref}[1]{\hyperref[#1]{Appendix~\ref*{#1}}}


\DeclareMathOperator{\E}{\mathbb{E}}     
\DeclareMathOperator*{\argmax}{arg\,max}

\newcommand{\DB}{\textbf{DB}}
\newcommand{\cand}{\mathcal{C}}

\newcommand{\stand}{\mathcal{S}}

\newcommand{\numcands}{C}
\newcommand{\altorder}{alt-order}

\newcommand{\cw}{1} 
\newcommand{\cx}{2} 
\newcommand{\cy}{3} 
\newcommand{\cz}{4} 

\title{Adaptively Weighted Audits of Instant-Runoff Voting Elections:
AWAIRE\thanks{Authors listed alphabetically.
Published in: Electronic Voting, E-Vote-ID 2023,
LNCS 14230, pp.\ 35--51, Springer, Cham (2023).
\url{https://doi.org/10.1007/978-3-031-43756-4_3}}}

\titlerunning{Adaptively Weighted Audits of IRV Elections: AWAIRE}

\author{
Alexander Ek       \inst{1}  \orcidID{0000-0002-8744-4805}  \and
Philip B. Stark    \inst{2}  \orcidID{0000-0002-3771-9604}  \and
Peter J. Stuckey   \inst{3}  \orcidID{0000-0003-2186-0459}  \and
Damjan Vukcevic    \inst{1}  \orcidID{0000-0001-7780-9586}
}

\authorrunning{Ek A, Stark PB, Stuckey PJ, Vukcevic D}

\institute{
Department of Econometrics and Business Statistics, Monash University, Clayton,
Australia
\and
Department of Statistics, University of California, Berkeley, CA, USA
\and
Department of Data Science and AI, Monash University, Clayton, Australia \\
\email{damjan.vukcevic@monash.edu}
}

\begin{document}

\maketitle

\begin{abstract}
An election audit is \emph{risk-limiting} if the audit limits (to a
pre-specified threshold) the chance that an erroneous electoral outcome
will be certified.  Extant methods for auditing instant-runoff voting
(IRV) elections are either not risk-limiting or require cast vote
records (CVRs), the voting system's electronic record of the votes on
each ballot.  CVRs are not always available, for instance, in
jurisdictions that tabulate IRV contests manually.

We develop an RLA method (AWAIRE) that uses adaptively weighted averages of
test supermartingales to efficiently audit IRV elections when CVRs are not
available.  The adaptive weighting `learns' an  efficient set of hypotheses
to test to confirm the election outcome.  When accurate CVRs are available,
AWAIRE can use them to increase the efficiency to match the performance of
existing methods that require CVRs.

We provide an open-source prototype implementation that can handle elections
with up to six candidates.  Simulations using data from real elections
show that AWAIRE is likely to be efficient in practice.  We discuss how
to extend the computational approach to handle elections with more
candidates.

Adaptively weighted averages of test supermartingales are a general tool,
useful beyond election audits to test collections of hypotheses
sequentially while rigorously controlling the familywise error rate.
\end{abstract}


\section{Introduction}

\emph{Ranked-choice} or \emph{preferential} elections allow voters to express
their relative preferences for some or all of the candidates, rather than
simply voting for one or more candidates.  Instant-runoff voting (IRV) is a
common form of ranked-choice voting.  IRV is used in political elections in
several countries, including all lower house elections in Australia.\footnote{%
Instant-runoff voting has been used in more than 500~political elections in
the~U.S. \url{https://fairvote.org/resources/data-on-rcv/} (accessed
18~July 2023).  It is also used by organisations; for instance, the `Best
Picture' Oscar is selected by instant runoff voting:
\url{https://www.pbs.org/newshour/arts/how-are-oscars-winners-decided-heres-how-the-voting-process-works}
(accessed 15~May 2023)}

A risk-limiting audit (RLA) is any procedure with a guaranteed minimum
probability of correcting the reported outcome if the reported outcome is
wrong.  RLAs never alter correct outcomes.  (\emph{Outcome} means the political
outcome---who won---not the particular vote tallies.) The \emph{risk limit}
$\alpha$ is the maximum chance that a wrong outcome will not be corrected.
Risk-limiting audits are legally mandated or authorised in approximately
15~U.S. states\footnote{%
See \url{https://www.ncsl.org/elections-and-campaigns/risk-limiting-audits}
(accessed 15~May 2023).}
and have been used internationally.  RAIRE \cite{blom2019raire} is the first
method for conducting RLAs for IRV contests.  RAIRE generates `assertions'
which, if true, imply that the reported winner really won.  Such assertions are
the basis of the SHANGRLA framework for RLAs \cite{shangrla}.

A \emph{cast vote record} (CVR) is the voting system's interpretation of the
votes on a ballot.  RAIRE uses CVRs to select the assertions to
test.\footnote{%
If the CVRs are linked to the corresponding ballot papers, then RAIRE can use
\emph{ballot-level comparison}, which increases efficiency.  See, e.g., Blom et
al.~\cite{blomEtal20}.}
Voting systems that tabulate votes electronically (e.g., using optical
scanners) typically generate CVRs, but in some jurisdictions (e.g., most lower
house elections in Australia) votes are tabulated manually, with no electronic
vote records.\footnote{%
IRV can be tabulated by hand, making piles of ballots with different
first-choices and redistributing the piles as candidates are eliminated, with
scrutineers checking that each step is followed correctly.}
Because RAIRE requires CVRs, it cannot be used to check manually tabulated
elections.  Moreover, while RAIRE generates a set of assertions that are
expected to be easy to check statistically if the CVRs are correct, if the CVRs
have a high error rate, then the assertions it generates may not hold
\emph{even if the reported winner actually won}, leading to an unnecessary full
hand count.

In this paper we develop an approach to auditing IRV elections that does not
require CVRs.  Instead, it adapts to the observed voter preferences as the
audit sample evolves, identifying a set of hypotheses that are efficient to
test statistically.  The approach has some statistical novelty and logical
complexity.  To help the reader track the gist of the approach, here is an
overview:
\begin{itemize}
\item Tabulating an IRV election results in a \emph{candidate elimination
    order}. A candidate elimination order that yields a winner other than the
    reported winner is an \emph{\altorder.} If there is sufficiently strong
    evidence that no \altorder{} is correct, we may safely conclude that
    the reported winner really won.
\item Each \altorder{} can be characterised by a set of \emph{requirements},
    necessary conditions for that elimination order to be correct.  If the data
    refute at least one requirement for each \altorder, the reported outcome is
    confirmed.
\item We construct a \emph{test supermartingale} for each requirement; a
    (predictable) convex combination of the test supermartingales for the
    requirements in an \altorder{} is a test supermartingale for that
    \altorder.
\item As the audit progresses, we update the convex combination for each
    \altorder{} to give more weight to the test supermartingales that are
    giving the strongest evidence that their corresponding requirements are
    false.
\item The audit has attained the risk limit $\alpha$ when the intersection test
    supermartingale for every \altorder{} exceeds $1/\alpha$ (or when every
    ballot has been inspected and the correct outcome is known).
\end{itemize}
The general strategy of adaptively re-weighting convex combinations of test
supermartingales gives powerful tests that rigorously control the sequential
familywise error rate.  It is applicable to a broad range of nonparametric and
parametric hypothesis testing problems.  We believe this is the first time
these ideas have been used in a real application.

To our knowledge, the SHANGRLA framework has until now been used to audit only
social choice functions for which correctness of the outcome is implied by
\emph{conjunctions} of assertions: if all the assertions are true, the contest
result is correct.  The approach presented here---controlling the familywise
error rate within groups of hypotheses and the per-comparison error rate across
such groups---allows SHANGRLA to be used to audit social choice functions for
which correctness is implied by \emph{disjunctions} of assertions as well as
conjunctions.  This fundamentally extends SHANGRLA.


\section{Auditing IRV contests}

We focus on IRV contests.  The set of candidates is $\cand$, with total number
of candidates $\numcands := |\cand|$.  A \emph{ballot} $b$ is an ordering of a
subset of candidates.  The number of ballots cast in the election is $B$.

Each ballot initially counts as a vote for the first-choice candidate on that
ballot.  The candidate with the fewest first-choice votes is eliminated (the
others remain `standing').  The ballots that ranked that candidate first are
now counted as if the eliminated candidate did not appear on the ballot: the
second choice becomes the first, etc.  This `eliminate the candidate with the
fewest votes and redistribute' continues until only one candidate remains
standing, the winner.  (If at any point there are no further choices of
candidate specified on a ballot, then the ballot is \emph{exhausted} and no
longer contributes any votes.) Tabulating the votes results in an
\emph{elimination order}: the order in which candidates are eliminated, with
the last candidate in the order being the winner.

\subsection{Alternative elimination orders}

In order to audit an IRV election we need to show that if any candidate other
than the reported winner actually won, the audit data would be `surprising,' in
the sense that we can reject (at significance level $\alpha$) the null
hypothesis that any other candidate won.

\begin{example}
\label{ex:four}
Consider a four-candidate election, with candidates $\cw$, $\cx$, $\cy$, $\cz$,
where $\cw$ is the reported winner.  We must be able to reject every
elimination order in which any candidate other than $\cw$ is eliminated last
(every \emph{\altorder}):
$[\cw,\cx,\cy,\cz]$, $[\cw,\cx,\cz,\cy]$, $[\cw,\cy,\cx,\cz]$,
$[\cw,\cy,\cz,\cx]$, $[\cw,\cz,\cx,\cy]$, $[\cw,\cz,\cx,\cy]$,
$[\cx,\cw,\cy,\cz]$, $[\cx,\cw,\cz,\cy]$, $[\cx,\cy,\cw,\cz]$,
$[\cx,\cz,\cw,\cy]$, $[\cy,\cw,\cx,\cz]$, $[\cy,\cw,\cz,\cx]$,
$[\cy,\cx,\cw,\cz]$, $[\cy,\cz,\cw,\cx]$, $[\cz,\cw,\cx,\cy]$,
$[\cz,\cw,\cy,\cx]$, $[\cz,\cx,\cw,\cy]$, $[\cz,\cy,\cw,\cx]$.
The other 6 elimination orders lead to $\cw$ winning: they are not
\altorder{s}.
\qed
\end{example}

To assess an \altorder, we construct \emph{requirements} that necessarily hold
if that \altorder{} is correct---then test whether those requirements hold.  If
one or more requirements for a given \altorder{} can be rejected statistically,
then that is evidence that the \altorder{} is not the correct elimination
order.  Blom et al.~\cite{blom2019raire} show that elimination orders can be
analysed using two kinds of statements, of which we use but one:\footnote{%
Blom et al.~\cite{blom2019raire} called these statements `IRV' rather than
`DB'.}
\medskip

\fbox{%
\parbox{0.9\textwidth}{%
`Directly Beats': $ \DB(i, j, \stand)$ holds if candidate~$i$ has more votes
than candidate~$j$, assuming that only the candidates $\stand \supseteq
\{i, j\}$ remain standing.  It implies that $i$ cannot be the next eliminated
candidate (since $j$ would be eliminated before $i$) if only the candidates
$\stand$ remain standing.
}
}

\subsection{Sequential testing using test supermartingales}
\label{sec:alpha}

Each requirement can be expressed as the hypothesis that the mean of a finite
list of bounded numbers is less than $1/2$.  Each such list results from
applying an \emph{assorter} (see Stark~\cite{shangrla}) to the preferences on
each ballot.  The assorters we use below all take values in $[0, 1]$.  For
example, consider the requirement $\DB(1, 2, \cand)$ that candidate~$1$ beats
candidate~$2$ on first preferences.  That corresponds to assigning a ballot the
value $1$ if it shows a first preference for candidate~$2$, the value $0$ if it
shows a first preference for $1$, and the value $1/2$ otherwise.  If the mean
of the resulting list of $B$ numbers is less than $1/2$, then the requirement
$\DB(1, 2, \cand)$ holds.

A stochastic process $(M_t)_{t \in \mathbb{N}}$ is a \emph{supermartingale}
with respect to another stochastic process $(X_t)_{t \in \mathbb{N}}$ if
$\E(M_t \mid X_1, \dots, X_{t-1}) \leqslant M_{t-1}$.  A \emph{test
supermartingale} for a hypothesis is a stochastic process that, if the null
hypothesis is true, is a nonnegative supermartingale with $M_0 := 1$.  By
Ville's inequality \cite{ville39}, which generalises Markov's inequality to
nonnegative supermartingales, the chance that a test supermartingale ever
exceeds $1/\alpha$ is at most $\alpha$ if the null hypothesis is true.  Hence,
we reject the null hypothesis if at some point $t$ we observe $M_t \geqslant
1/\alpha$.  The maximum chance of the rejection being in error is $\alpha$.

Let $X_1, X_2, \dots$ be the result of applying the assorter for a particular
requirement to the votes on ballots drawn sequentially at random without
replacement from all of the $B$ cast ballots.  We test the requirement using
the ALPHA test supermartingale for the hypothesis that the mean of the $B$
values of the assorter is at most $\mu_0$ is
\[
M_j = \prod_{i=1}^j \left(\frac{X_i}{\mu_i} \cdot
                          \frac{\eta_i - \mu_i }{1 - \mu_i} +
                          \frac{ 1     - \eta_i}{1 - \mu_i}\right),
    \quad j = 1, 2, \dots, B,
\]
where
\[
\mu_j = \frac{B\mu_0 - \sum_{i=1}^{j-1} X_i}{B - j + 1}
\]
is the mean of the population just before the $j$th ballot is drawn (and is
thus the value of $\E X_j$) if the null hypothesis is true.
The value of $M_j$ decreases monotonically in $\mu_0$, so it suffices to
consider the largest value of $\mu_0$ in the null hypothesis, i.e., $\mu_0 =
1/2$ \cite{stark2023alpha}.
The value $\eta_j$ can be thought of as a (possibly biased) estimate of the
true assorter mean for the ballots remaining in the population just before the
$j$th ballot is drawn.  We use the `truncated shrinkage' estimator suggested by
Stark~\cite{stark2023alpha}:
\[
\eta_j =
    \min\left[
    \max\left(
        \frac{d\eta_0 + \sum_{i=1}^{j-1} X_i}{d+j-1},\,
        \mu_j + \epsilon_j\right),\,
        1                 \right] .
\]
The parameters $\epsilon_j = (\eta_0 - \mu)/(2\sqrt{d+j-1})$ form a nonnegative
decreasing sequence with $\mu_j < \eta_j \leqslant 1$.
The parameters $\eta_0$ and $d$ are tuning parameters.
The ALPHA supermartingales span the family of \emph{betting} supermartingales,
discussed by \cite{waudby-smith2023betting}: setting $\eta_j$ in ALPHA is
equivalent to setting $\lambda_j$ in betting supermartingales
\cite{stark2023alpha}.


\section{Auditing via adaptive weighting (AWAIRE)}
\label{sec:methods}

\subsection{Eliminating elimination orders using `requirements'}

We can formulate auditing an IRV contest as a collection of hypothesis tests.
To show that the reported winner really won, we consider every elimination
order that would produce a different winner (every \altorder).
The audit stops without a full hand count if it provides sufficiently strong
evidence that no \altorder{} occurred.
Suppose there are $m$ \altorder{s}.
Let $H_0^i$ denote the hypothesis that \altorder{} $i$ is the true elimination
order, $i = 1, \dots, m$.
These partition the global null hypothesis,
\[
    H_0 = H_0^1 \cup \dots \cup H_0^m.
\]
If we reject all the null hypotheses $H_0^1, \dots, H_0^m$, then we have also
rejected $H_0$ and can certify the outcome of the election.

For each \altorder{} $i$, we have a set of \emph{requirements} $R_i = \{R_i^1,
R_i^2, \dots , R_i^{r_i}\}$ that necessarily hold if $i$ is the true
elimination order, i.e.,
\[
    H_0^i \subseteq R_i^1 \cap R_i^2 \cap \dots \cap R_i^{r_i}.
\]
If any of these requirements is false, then \altorder{} $i$ is not the true
elimination order.
If
\[
    H_0^i = R_i^1 \cap R_i^2 \cap \dots \cap R_i^{r_i}
\]
then $R_i$ is a \emph{complete} set of requirements: they are necessary and
sufficient for elimination order $i$ to be correct.
One way to create a complete set is to take all \DB{} requirements that
completely determine each elimination in the given elimination order.

\begin{example}
A complete set of requirements for the elimination order $[\cw,\cx,\cy,\cz]$
is:
  $\DB(\cz,\cy,\{\cy,\cz\})$,
  $\DB(\cz,\cx,\{\cx,\cy,\cz\})$,
  $\DB(\cy,\cx,\{\cx,\cy,\cz\})$,
  $\DB(\cz,\cw,\{\cw,\cx,\cy,\cz\})$,
  $\DB(\cy,\cw,\{\cw,\cx,\cy,\cz\})$, and
  $\DB(\cx,\cw,\{\cw,\cx,\cy,\cz\})$.
If we reject any of these, then we can reject the elimination order
$[\cw,\cx,\cy,\cz]$.
\qed
\end{example}

We can rule out \altorder{} $i$ by rejecting the intersection hypothesis $R_i^1
\cap \dots \cap R_i^{r_i}$.  The test supermartingales for the individual
requirements are dependent because all are based on the same random sample of
ballots.  \autoref{sec:adaptive-weighting} shows how to test the intersection
hypothesis, taking into account the dependence.

\subsubsection{`Requirements' vs `assertions'.}

SHANGRLA \cite{shangrla} uses the term `assertions.' Requirements and
assertions are statistical hypotheses about means of assorters applied to the
votes on all the ballots cast in the election.  `Assertions' are hypotheses
whose conjunction is \emph{sufficient} to show that the reported winner really
won: if all the assertions are true, the reported winner really won.
`Requirements' are hypotheses that are \emph{necessary} if the reported winner
really lost---in a particular way, e.g., because a particular \altorder{}
occurred.  Loosely speaking, assertions are statements that, if true, allow the
audit to stop; while requirements are statements that, if false, allow the
audit to stop.  To stop without a full hand count, an assertion-based audit
needs to show that every assertion is true.  In contrast, a requirement-based
audit needs to show that at least one requirement is false in each element
$H_0^i$ of a partition of the null hypothesis $H_0 = \cup_i H_0^i$.  (In
AWAIRE, the partition corresponds to the \altorder{s}.)

\subsection{Adaptively weighted test supermartingales}
\label{sec:adaptive-weighting}

Given the sequentially observed ballots, we can construct a test
supermartingale (such as ALPHA) for any particular requirement.  To test a
given hypothesis $H_0^i$, we need to test the intersection of the requirements
in the set $R_i$.  We now describe how we test that intersection hypothesis,
despite the dependence among the test supermartingales for the separate
requirements.  The test involves forming weighted combinations of the terms in
the test supermartingales for individual requirements in such a way that the
resulting process is itself a test supermartingale for the intersection
hypothesis.  This is somewhat similar to the methods of combining test
supermartingales described by Vovk \& Wang~\cite{vovkWang21}.

The quantities defined in this section, such as $E_{r,t}$, are for a given set
of requirements $R_i$ and thus implicitly depend on $i$.  For brevity, we omit
$i$ in the notation.

At each time $t$, a ballot is drawn without replacement, and the assorter
corresponding to each $R_i^r$, $r = 1, \dots, r_i$ is computed, producing the
values $X_t^r$, $r = 1, \dots, r_i$.  Let $(E_{r,t})$ be the test
supermartingale for requirement $r$.  The test supermartingale can be written
as a telescoping product:\footnote{%
This is always possible, by taking $e_{r,t} := E_{r,t} / E_{r,t-1}$.}
\[
    E_{r,t} := \prod_{k=0}^{t} e_{r,k},
\]
with $E_{r,0} := 1$ for all $r$ and
\begin{equation}
\label{eq:e-expect}
    \mathbb{E}(e_{r,k} \mid (X_\ell^r)_{\ell=0}^{k-1}) \leqslant 1,
\end{equation}
where the conditional expectation is computed under the hypothesis that
requirement $r$ is true.
(This last condition amounts to the supermartingale property.)
We refer to these as \emph{base} test supermartingales.

For each $k$, let $\{w_{r,k}\}_{r=1}^{r_i}$ be nonnegative \emph{predictable}
numbers: $w_{r,t}$ can depend on the values $\{X_k^r\}$, $r = 1, \dots, r_i$,
$k = 0, \dots, t - 1$ but not on data collected on or after time $t$.  Define
the stochastic process formed by multiplying convex combinations of terms from
the base test supermartingales using those weights:
\[
    E_t := \prod_{k=1}^t \frac{\sum_{r=1}^{r_i} w_{r,k} e_{r,k}}
                              {\sum_{r=1}^{r_i} w_{r,k}},
    \quad t = 0, 1, \dots,
\]
with $E_0 := 1$.
This process, which we call an \emph{intersection} test supermartingale,
is a test supermartingale for the intersection of the $r_i$ hypotheses:
Clearly $E_t \geqslant 0$ and $E_0 :=1$, and
if all the hypotheses are true,
\begin{eqnarray*}
\E\left(E_t  \mid (X_k^r)_{k=1}^{t-1}, r=1, \dots, r_i \right) &=&
\E\left(E_{t-1} \frac{\sum_{r=1}^{r_i} w_{r,k} e_{r,k}}
                     {\sum_{r=1}^{r_i} w_{r,k}} \mid
        (X_k^r)_{k=1}^{t-1}, r=1, \dots, r_i \right)  \\
&=& E_{t-1} \E\left(\frac{\sum_{r=1}^{r_i} w_{r,k} e_{r,k}}
                         {\sum_{r=1}^{r_i} w_{r,k}} \mid
                    (X_k^r)_{k=1}^{t-1}, r=1, \dots, r_i \right)  \\
&=& E_{t-1} \frac{\sum_{r=1}^{r_i} w_{r,k} \;
                  \E\left(e_{r,k} \mid (X_k^r)_{k=1}^{t-1} \right)}
                 {\sum_{r=1}^{r_i} w_{r,k}}  \\
&\leqslant&
    E_{t-1} \frac{\sum_{r=1}^{r_i} w_{r,k}}
                 {\sum_{r=1}^{r_i} w_{r,k}}
 =  E_{t-1},
\end{eqnarray*}
where the penultimate step follows from \autoref{eq:e-expect}.
Thus $(E_t)$ is a test supermartingale for the intersection of the
requirements.  The base test supermartingale for any requirement that is false
is expected to grow in the long run (the growth rate depends on the true
assorter values and the choice of base test supermartingales).  We aim to make
$E_t$ grow as quickly as the fastest-growing base supermartingale by giving
more weight to the terms from the base supermartingales that are growing
fastest.

For example, we could take the weights to be proportional to the base values in
the previous timestep, $w_{r,t} = E_{r,t-1}$.  More generally, we can explore
other functions of those previous values, see below for some options.  Unless
stated otherwise, we set the initial weights for the requirement to be equal.

This describes how we test an individual \altorder.
The same procedure is used in parallel for every \altorder.
Because the audit stops without a full handcount only if \emph{every}
\altorder{} is ruled out, there is no multiplicity issue.

\subsubsection{Setting the weights.}
\label{sec:weights}

We explored three ways of picking the weights:
\begin{description}
\item[Linear.]  Proportional to previous value, $w_{r,t} := E_{r,t-1}$.
\item[Quadratic.]  Proportional to the square of the previous value,
                                                $w_{r,t} := E_{r,t-1}^2$.
\item[Largest.]  Take only the largest base supermartingale(s) and ignore the
    rest, $w_{r,t} := 1$ if $r \in \argmax_{r'} E_{r',t-1}$; otherwise,
    $w_{r,t} := 0$.
\end{description}

\subsubsection{Using ALPHA with AWAIRE.}

The adaptive weighting scheme described above can work with any test
supermartingales.  In our implementation, we use ALPHA with the truncated
shrinkage estimator to select $\eta_t$ (see \autoref{sec:alpha}); it would be
interesting to study the performance of other test supermartingales, for
instance, some that use the betting strategies in Waudby-Smith \&
Ramdas~\cite{waudby-smith2023betting}.

In our experiments (see \autoref{sec:results}), the intersection test
supermartingales were evaluated after observing each ballot.  However, for
practical reasons, we updated the weights only after observing every 25 ballots
rather than every ballot; this does not affect the validity (the risk limit is
maintained), only the adaptivity.  Initial experiments seem to indicate that
updating the weights more frequently often slightly favours lower sample sizes,
but not always.

\subsubsection{Using CVRs.}

If accurate CVRs are available, then we can use them to `tune' AWAIRE and ALPHA
to be more efficient for auditing the given contest.  We explore several
options in \autoref{sec:using-cvrs}.  If CVRs are available and are `linked' to
the paper ballots in such a way that the CVR for each ballot card can be
identified, AWAIRE can also be used with a ballot-level comparison audit, which
could substantially reduce sample sizes compared to ballot-polling.  See, e.g.,
Stark~\cite{shangrla}.  We have not yet studied the performance of AWAIRE for
ballot-level comparison audits, only ballot-polling audits.


\section{Analyses and results}
\label{sec:results}

To explore the performance of AWAIRE, we simulated ballot-polling audits using
a combination of real and synthetic data (see below).  Each sampling experiment
was repeated for 1,000 random permutations of the ballots, each corresponding
to a sampling order (without replacement).  For each contest, the same 1,000
permutations were used for every combination of tests and tuneable parameters.

In each experiment, sampling continued until either the method confirmed the
outcome or every ballot had been inspected.  We report the mean sample size
(across the 1,000 permutations) for each method.

The ballots were selected one at a time without replacement, and the base test
supermartingales were updated accordingly.  However, to allow the experiments
to complete in a reasonable time, we only updated the weights after every 25
ballots were sampled.  This is likely to slightly inflate the required sample
sizes due to the reduced adaptation.

We repeated all of our analyses with a risk limit of 0.01, 0.05, 0.1, and 0.25.
The results were qualitatively similar across all choices, therefore we only
show the results for $\alpha = 0.01$.

\subsection{Data and software}

We used data from the New South Wales (NSW) 2015 Legislative Assembly election
in Australia.\footnote{Source: \url{https://github.com/michelleblom/margin-irv}
(accessed 17~April 2023)}  We took only the 71 contests with 6 or fewer
candidates (due to computational constraints: future software will support
elections with more candidates).  The contests each included about 40k--50k
ballots. Our software implementation of AWAIRE is publicly
available.\footnote{\url{https://github.com/aekh/AWAIRE}}

We supplemented these data with 3 synthetic `pathological' contests that were
designed to be difficult to audit, using the same scheme as Everest et
al.~\cite{everest2023dtree}.  Each contest had 6 candidates and 56k ballots,
constructed as follows.  Candidates: $a$ the (true) winner, $b$ an alternate
winner, and candidates $c_1, c_2, c_3, c_4$.  Ballots:
\begin{itemize}
\item $16000 + 2m$ ballots of the form $[a]$,
\item $8000 - 2m$ ballots of the form $[b]$,
\item 8000 ballots of the form $[c_i, b, a]$ for each $i \in \{1,2,3,4\}$.
\end{itemize}
We used $m \in \{2.5, 25, 250\}$ to define the 3 pathological contests.

In each of these contests, $b$ is eliminated first, then each of the $c_i$ is
eliminated, making $a$ the winner.  If any $c_i$ is eliminated first by mistake
(e.g., due to small errors in the count), then $b$ does not get eliminated and
instead will collect all of the votes after each elimination and become the
winner.  A random sample of ballots, such as is used in an audit, will likely
often imply the wrong winner.

We calculated the \emph{margin} of each contest using
\texttt{margin-irv}~\cite{blom2018margin} to allow for easier interpretation of
the results.  The margin is the minimum number of ballots that need to be
changed so that the reported winner is no longer the winner, given they were
the true winner originally.  For easier comparison across contests, we report
the margin as proportion of the total ballots rather than as a count.

\subsection{Comparison of weighting schemes}

We tested AWAIRE with the three weighting schemes described earlier (Linear,
Quadratic, and Largest).  For the test supermartingales, we used ALPHA with
$\eta_0 = 0.52$ and $d = 50$.  For each simulation, we first set the true
winner to be the reported winner, and then repeated it with the closest
runner-up candidate (based on the margin) as the reported winner.  This allowed
us to explore scenarios where the reported winner was false, in order to verify
the risk limit (in all cases, the proportion of such simulations the led to
certifying the wrong winner was lower than the risk limit).

\begin{figure}[t]
\centering
\includegraphics[width=\textwidth]{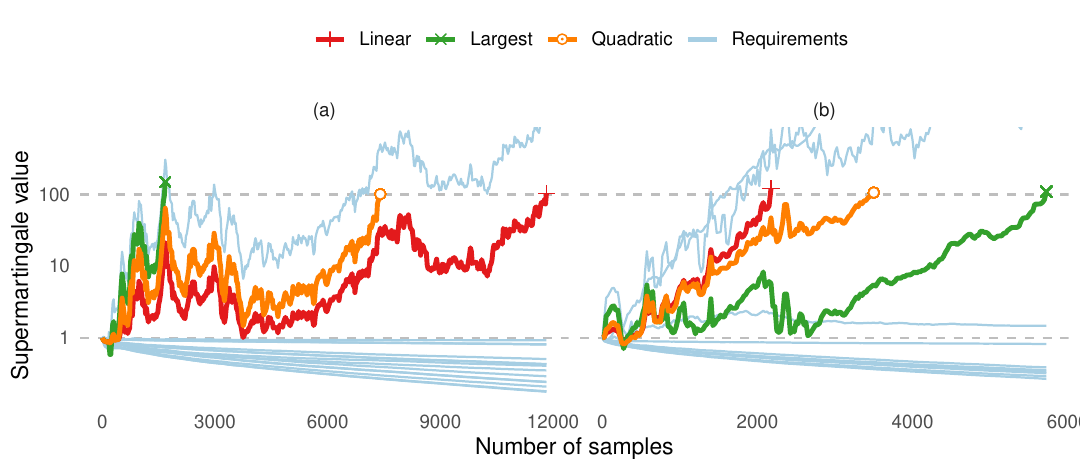}
\caption{\textbf{Examples of test supermartingales for a set of requirements.}
The plots show how our test supermartingales evolved as we sampled increasingly
more ballots in a particular audit with risk limit $\alpha = 0.01$.  Each
panel refers to a particular null hypothesis (i.e., a single \altorder{})
for a particular contest.  The lines in light blue show the test
supermartingales for each requirement used to test that hypothesis; the
bold lines show our adaptively weighted combination across all requirements
(using the weighting schemes as indicated by colour).
Panel~(a): NSW 2015 Upper Hunter contest (true elimination order
    $[1, 2, 3, 0, 4, 5]$) with a hypothesised order $[1, 2, 3, 0, 5, 4]$.
Panel~(b): NSW 2015 Prospect contest (true elimination order
    $[3, 0, 4, 1, 2]$) with a hypothesised order $[0, 3, 4, 2, 1]$.
The horizontal lines indicate the start (1) and target ($1 / \alpha = 100$)
values; we stop sampling and reject the null hypothesis when the
intersection test supermartingale exceeds the target value.}
\label{fig:martingale-examples}
\end{figure}

\autoref{fig:martingale-examples} illustrates an example of how the test
supermartingales evolved in two simulations.  Panel~(a) is a more typical
scenario, while panel~(b) is an illustration of the rare scenarios where
Largest is worst (due to competing and `wiggly' base supermartingales).

\begin{figure}[t]
\centering
\includegraphics[width=\textwidth]{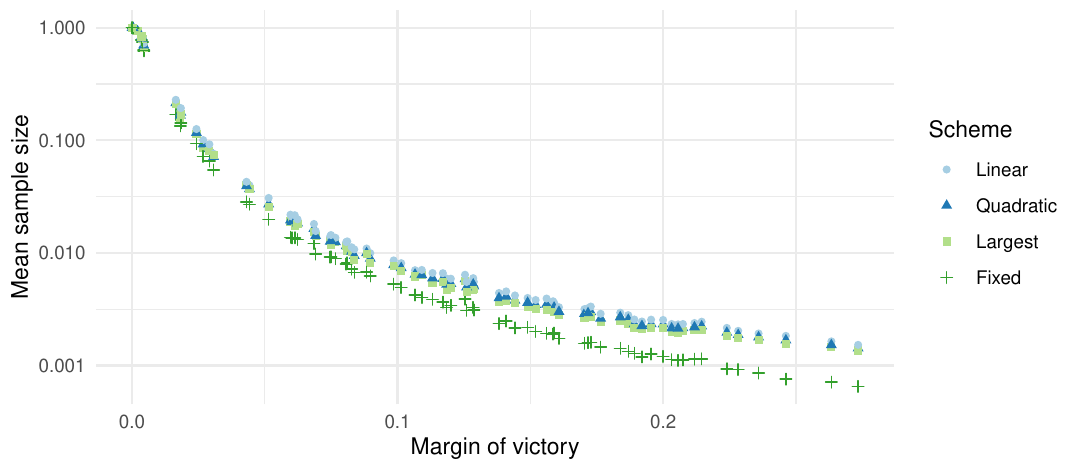}
\caption{\textbf{Comparison of weighting schemes.}
The mean sample size (across 1,000 simulations; shown on a log scale) versus
the margin, both shown as a proportion out of the total ballots in a
contest.  Each point depicts a single contest and weighting scheme, the
latter distinguished by colour and point type as indicated.  The `Fixed'
scheme is only shown for reference: adaptive weighting was disabled and
only the best requirements were used.
}
\label{fig:compare-weights}
\end{figure}

\autoref{fig:compare-weights} summarises the performance of the different
weighting schemes across a large set of contests.  Some more details for a
selected subset of contests are shown in top part of \autoref{tab:results}.

The three weighting schemes differ in how `aggressively' they favour the
best-looking requirements at each time point.  In our experiments, the more
aggressive schemes consistently performed better, with the Largest scheme
achieving the best (lowest) mean sample sizes.  On this basis, and the
simplicity of the Largest scheme, we only used this scheme for the later
analyses.

A key feature of AWAIRE is that it uses the observed ballots to `learn' which
requirements are the easiest to reject for each elimination order and adapts
the weights throughout the audit to take advantage.  To assess the statistical
`cost' of the learning, we also ran simulations that used a fixed weight of 1
for the test supermartingales for the requirements that proved easiest to
reject, and gave zero weight to the other requirements (we call this the
`Fixed' scheme).\footnote{%
This is equivalent to a scenario where we have fully accurate CVRs available
and decide to keep weights fixed.  We explore such options in the next
section.}
The performance in this mode is shown in \autoref{fig:compare-weights} as green
crosses. The Fixed version gave smaller mean sample sizes, getting as small as
55\% of the Largest.  This shows that adaptation less than doubles the sample
size.

\begin{table}[t]
\centering
\caption{\textbf{Selected results.}
The mean sample size from experiments using a risk limit of $\alpha = 0.01$,
across a subset of contests, in 1,000 replications.  The contest margins
range from very close (Lismore) to a very wide margin (Castle Hill).  The
top part of the table shows results from analyses that did not use CVRs;
the bottom part shows results from analyses using CVRs without errors.  The
column labeled $d$ is the value of the ALPHA $d$ parameter.}
\begin{tabular}{llrrrrrrr}
\toprule
\multicolumn{3}{r}{\textbf{Contest:}}         & Lismore & Monaro & Auburn & Maroubra & Cessnock & Castle Hill \\
\multicolumn{3}{r}{\textbf{No.\ candidates:}} & 6      & 5      & 6      & 5      & 5      & 5      \\
\multicolumn{3}{r}{\textbf{Margin:}}          & 0.44\% & 2.43\% & 5.15\% & 10.1\% & 20.0\% & 27.3\% \\
\multicolumn{3}{r}{\textbf{Total ballots:}}   & 47,208 & 46,236 & 44,011 & 46,533 & 45,942 & 48,138 \\
\addlinespace
\textbf{Method} & \textbf{Weights} & \multicolumn{1}{c}{$d$} &
\multicolumn{6}{c}{\textbf{Mean sample size}}  \\
\cmidrule{1-3}
\cmidrule(l){4-9}
\multicolumn{3}{l}{\footnotesize\textsl{No CVRs}}  \\
AWAIRE
 & Linear    & 50 &  34,246  &  5,822  &  1,354  &  378  &  117  &  73  \\
 & Quadratic & 50 &  32,988  &  5,405  &  1,195  &  343  &  107  &  69  \\
 & Largest   & 50 &  32,534  &  5,217  &  1,130  &  320  &   98  &  65  \\
\addlinespace
\multicolumn{3}{l}{\footnotesize\textsl{With error-free CVRs}}  \\
AWAIRE
 & Largest   &  50 &  32,312  &  5,172  &  1,074  &  283  &  60  &  33  \\
 & Largest   & 500 &  31,790  &  4,458  &    942  &  265  &  59  &  33  \\
 & Fixed     &  50 &  29,969  &  4,317  &    876  &  230  &  55  &  31  \\
 & Fixed     & 500 &  29,756  &  3,912  &    781  &  212  &  54  &  31  \\
RAIRE
 & ---       &  50 &  31,371  &  4,260  &    876  &  230  &  56  &  34  \\
 & ---       & 500 &  31,034  &  3,862  &    781  &  212  &  54  &  33  \\
\bottomrule
\end{tabular}
\label{tab:results}
\end{table}

\subsection{Using CVRs (without errors)}
\label{sec:using-cvrs}

We compare AWAIRE to RAIRE \cite{blom2019raire}, the only other extant RLA
method for IRV contests.  Since RAIRE requires CVRs, we considered several ways
in which we could use AWAIRE when CVRs are available.  We explored choices for
the following:
\begin{description}
\item[Starting weights.]  Using the CVRs we can calculate the (reported) margin
    for each requirement, allowing us to determine the easiest requirement to
    reject for each null hypothesis (assuming the CVRs are accurate).  We
    gave each such requirement a starting weight of 1, and the other
    requirements a starting weight of~0.  Other choices are possible (e.g.,
    weights set according to some function of the margins) but we did not
    explore them.
\item[Weighting scheme.]  If the CVRs are accurate, then it would be optimal to
    keep the starting weights fixed across time (similar to RAIRE).
    Alternatively, we can allow the weights to adapt as usual to the
    observed ballots, in case the CVRs are inaccurate.  We explored both
    choices, using only the Largest weighting scheme (which performed best
    in our comparison, above).
\item[Test supermartingales.]  Having CVRs available allows us to tune ALPHA
    for each requirement by setting $\eta_0$ to the reported assorter mean
    (based on the CVRs).  We allowed ALPHA to adapt by setting $d = 500$
    (adapt slowly) or $d = 50$ (adapt quickly).  For any requirements that
    the CVRs claim are true (i.e., consistent with the null hypothesis,
    with the assorter mean at most 0.5), we used a default value of
    $\eta_0 = 0.52$.
\end{description}
For comparison, we ran RAIRE with the same set of choices for the test
supermartingales.  For this analysis, we used accurate CVRs (no errors), the
best-case scenario for RAIRE and for any choices where adaptation is slow or
`switched off' (such as keeping the weights fixed).

\begin{figure}[t]
\centering
\includegraphics[width=\textwidth]{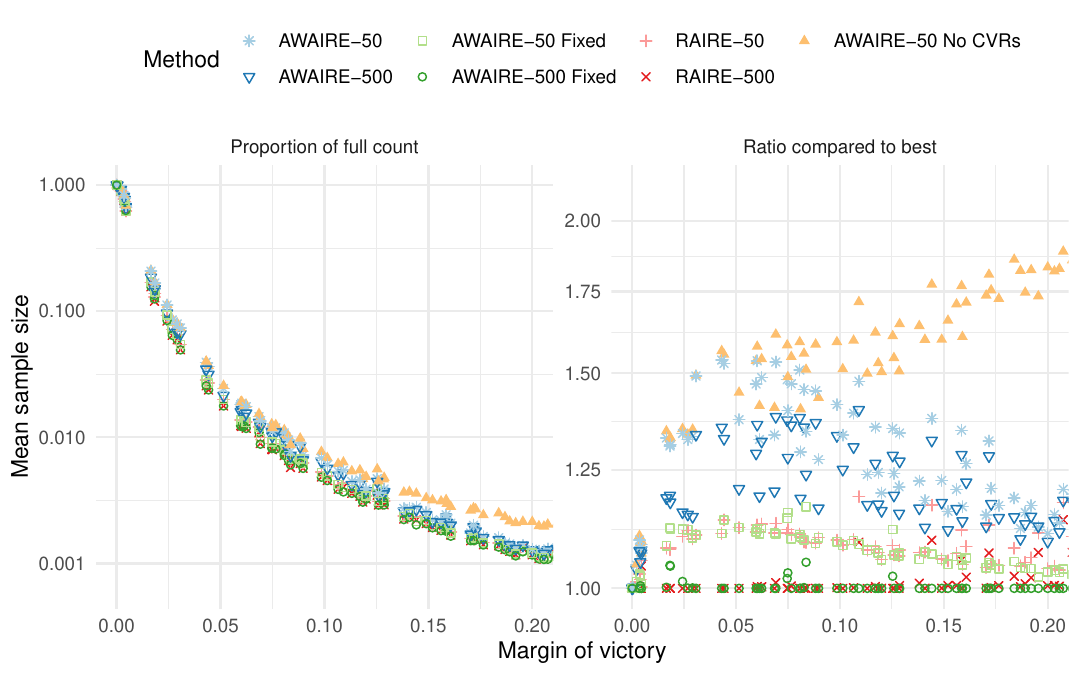}
\caption{\textbf{Comparison of methods when using accurate CVRs.}
Left panel: similar to \autoref{fig:compare-weights} but now showing different
    variants of AWAIRE all using the Largest scheme, and RAIRE.
Right panel: showing mean sample size as a ratio compared to the best method
    for each contest.
`Fixed' means the weights were kept fixed throughout the audit.
`No CVRs' means AWAIRE was not provided the CVRs to set the starting weights.
    The numbers 50/500 specify the value $d$ used for ALPHA.}
\label{fig:performance-using-cvrs}
\end{figure}

\autoref{fig:performance-using-cvrs} summarises the results, with a selected
subset shown in the bottom part of \autoref{tab:results}.
RAIRE and AWAIRE Fixed are on par when the CVRs are perfectly accurate, with
both methods being equal most of the time.
For margins up to $10\%$, RAIRE is ahead (albeit slightly) more often than
AWAIRE Fixed is;
for margins above $10\%$, AWAIRE Fixed is instead more often slightly ahead.

For both AWAIRE and RAIRE, the `less adaptive' versions performed better than
their `more adaptive' versions (there is no need to adapt if there are no
errors).  The largest ratio between the best setup and `AWAIRE-50 No CVRs' is
2.14, which occurs around a margin of 27.3\%.  However, at that margin, it
translates to a difference of less than 35 ballots.

Interestingly, the difference between the various versions of AWAIRE is small.
Across the different margins, they maintain the relative order from the least
informed (No CVRs) to the most informed and least adaptive (Fixed weights,
$d = 500$).
The cost of non-information in terms of mean sample size is surprising low,
particularly when the margin of victory is small: there is little difference
between `AWAIRE-50 No CVRs' and `AWAIRE-50'.  As the margin grows, the relative
difference becomes more substantial but the ratio never exceeds 1.97, and at
this stage the absolute difference is small (within 50 ballots).

\autoref{tab:results} gives more detail on a set of elections. For the smallest
margin election, AWAIRE Fixed using CVRs outperforms RAIRE, which outperforms
AWAIRE Largest using CVRs, which outperforms AWAIRE without CVRs; but the
relative difference in the number of ballots required to verify the result is
small (about 14\%).  In this case, the variants of AWAIRE have similar
workloads, with or without CVRs.  For larger margins ($>$ 5\%), the auditing
effort falls, and the relative differences between AWAIRE and RAIRE become
negligible.

Overall, while AWAIRE with no CVRs can require much more auditing effort than
when perfect CVRs are available, for small margins the relative cost difference
is small, and for larger margins the absolute cost difference is small.  This
shows that AWAIRE is certainly a practical approach to auditing IRV elections
without the need for CVRs (if doing a ballot-polling audit).

\subsection{Using CVRs with permuted candidate labels}
\label{sec:using-cvrs-with-errors}

We sought to repeat the previous comparison but with errors introduced into the
CVRs.  There are many possible types of errors and, as far as we are aware, no
existing large dataset from which we could construct a realistic error model.
A thorough analysis of possible error models is beyond the scope of this paper.
For illustrative purposes, we explored scenarios where the candidate labels are
permuted in the CVRs, the same strategy adopted by Everest et
al.~\cite{everest2023dtree}.

While this type of error can plausibly occur in practice, we use it here for
convenience: it allows us to easily generate scenarios where the reported
winner is correct but the elimination order implied by the CVRs is incorrect.
This is likely to lead RAIRE to escalate to a full count if it selects a
suboptimal choice of assertions.  We wanted to see whether in such scenarios
AWAIRE could `recover' from a poor starting choice by taking advantage of
adaptive weighting.

\begin{table}[t]
\centering
\caption{\textbf{Comparison of methods when using CVRs with errors.}
Mean sample sizes for experiments using the
NSW 2015 Strathfield contest (46,644 ballots, 1.65\% margin)
and CVRs with different permutations of the candidate labels (leading
to different reported elimination orders).
The columns refer to groups of one or more permutations for which we observed
largely similar results for each of the auditing methods; the corresponding
mean sample sizes reported in the table were the average across the
permutations in the group (`all' = 46,644).
The true elimination order is $[1,2,3,4,5]$.
Notation for reported elimination orders:
an integer means the given candidate is in that place in the order,
a crossed-out integer means the given candidate is \emph{not} in that place,
a dot ($\cdot$) means any unmentioned candidate can be in that place, and
the final column includes all orders with incorrect winners.
}
\begin{tabular}{lrrrrrrr}
\toprule
                & \multicolumn{7}{c}{\textbf{Reported elimination order}} \\
\cmidrule{2-8}
\textbf{Method} &
$\left[\begin{array}{c} 1 \\ 2 \\ 3 \\ 4 \\ 5 \end{array}\right]$ &
$\left[\begin{array}{c} 2 \\ 1 \\ 3 \\ 4 \\ 5 \end{array}\right]$ &
$\left[\begin{array}{c} \cdot \\ \cdot \\ \cancel{3} \\ 4 \\ 5 \end{array}\right]$ &
$\left[\begin{array}{c} \cdot \\ \cdot \\ 4 \\ \cdot \\ 5 \end{array}\right]$ &
$\left[\begin{array}{c} \cdot \\ 4 \\ \cdot \\ \cdot \\ 5 \end{array}\right]$ &
$\left[\begin{array}{c} 4 \\ \cdot \\ \cdot \\ \cdot \\ 5 \end{array}\right]$ &
~\rotatebox[origin=c]{90}{Other}~ \\
\midrule
AWAIRE-50 No CVRs &   9,821  &   9,821  &   9,821  &   9,821  &   9,821  &   9,821  &     all  \\
AWAIRE-50         &   9,694  &   9,717  &   9,810  &  14,229  &  15,714  &  15,929  &     all  \\
AWAIRE-500        &   8,656  &   8,863  &   9,052  &  25,410  &  29,274  &  29,786  &     all  \\
AWAIRE-50 Fixed   &   7,912  &   7,914  &     all  &  46,462  &     all  &     all  &     all  \\
AWAIRE-500 Fixed  &   7,315  &   7,315  &     all  &  46,460  &     all  &     all  &     all  \\
RAIRE-50          &   7,875  &   7,875  &   7,875  &  46,504  &  46,504  &  46,504  &  46,621  \\
RAIRE-500         &   7,301  &   7,301  &   7,301  &  46,318  &  46,272  &  46,225  &  46,621  \\
\bottomrule
\end{tabular}
\label{tab:performance-using-cvrs-with-errors}
\end{table}

We simulated audits for a particular 5-candidate contest, exploring all $5! =
120$ possible permutations of the candidate labels in the CVRs.  The results
are summarised in \autoref{tab:performance-using-cvrs-with-errors}.
Without label permutation, the results were consistent with
\autoref{sec:using-cvrs}.  Swapping the first two eliminated candidates made
little difference.  Permuting the first three eliminated candidates exposed the
weakness of the Fixed strategies, which nearly always escalated to full counts.
When the runner-up candidate was moved to be reportedly eliminated earlier in
the count, RAIRE nearly always escalated to a full count, but AWAIRE performed
substantially better (at least for $d = 50$), demonstrating AWAIRE's ability to
`recover' from CVR errors.  For permutations where the reported winner was
incorrect, AWAIRE always led to full count, while RAIRE incorrectly certified
0.3\% of the time.

\enlargethispage{\baselineskip}


\section{Discussion}

AWAIRE is the first RLA method for IRV elections that does not require CVRs.
AWAIRE may be useful even when CVRs are available, because it may avoid a full
handcount when the elimination order implied by the CVRs is wrong but the
reported winner really won---a situation in which RAIRE is likely to lead to an
unnecessary full handcount.

Comparisons of AWAIRE workloads with and without the adaptive weighting shows
that the `cost' of this feature is relatively small (i.e., how many extra
samples are required when `learning', compared to not having to do any
learning).  However, we also saw a sizable difference in performance between
AWAIRE with adaptive weighting and methods that had both access to and complete
faith in (correct) CVRs (i.e., RAIRE and AWAIRE Fixed).

In some scenarios, RAIRE was slightly more efficient than AWAIRE (similarly
configured).  The two main differences between these methods are (i)~RAIRE uses
an optimisation heuristic to select its assertions and (ii)~RAIRE has a richer
`vocabulary' of assertions to work with than the current form of AWAIRE, which
only considers \DB{} for alternate candidate elimination orders.  AWAIRE can be
extended to use additional requirements, similar to the \textbf{WO} assertions
of Blom et al.~\cite{blomEtal20} (which asserts that one candidate always gets
more votes initially than another candidate ever gets).  Rejecting one such
assertion can rule out many \altorder{s}.  Adding requirements to AWAIRE that
are similar to these assertions may reduce the auditing effort, since they are
often easy to reject.

Our current software implementation becomes inefficient when there are many
candidates, because the number of null hypotheses we need to reject is
factorial in the number of candidates $\numcands$, and the number of \DB{}
requirements we need to track is $O(\numcands!\,\numcands^2)$.  Future work
will investigate a \emph{lazy} version of AWAIRE, where rather than consider
all requirements for all \altorder{s}, we only consider a limited set of
requirements (e.g., only those concerning the last 2 remaining candidates).
Once we have rejected many \altorder{s} with these few requirements, which we
are likely to do early on, we can then consider further requirements for the
remaining \altorder{s} (e.g., concerning the last 3~candidates).  Again, once
even more \altorder{s} have been rejected, with the help of these newly
introduced requirements, we can then consider the last 4 remaining candidates,
and so on.  This lazy expansion process should result in considering far fewer
than the $O(\numcands!\,\numcands^2)$ \DB{} requirements in all.

This work extends SHANGRLA in a fundamental way, allowing it to test
disjunctions of assertions, not just conjunctions.  The adaptive weighting
scheme we develop using convex combinations of test supermartingales is quite
general; it solves a broad range of statistical problems that involve
sequentially testing intersections and unions of hypotheses using dependent or
independent observations.

\enlargethispage{\baselineskip}


\subsubsection{Acknowledgements.}

We thank Michelle Blom, Ronald Rivest and Vanessa Teague for helpful
discussions and suggestions.
This work was supported by the Australian Research Council
(Discovery Project DP220101012, OPTIMA ITTC IC200100009)
and the U.S.\ National Science Foundation (SaTC~2228884).


\bibliographystyle{splncs04}
\bibliography{references}


\end{document}